\documentclass[runningheads]{llncs}
\usepackage[T1]{fontenc}
\usepackage{graphicx}
\usepackage{hyperref}
\usepackage{amsmath}        
\usepackage{breqn}
\usepackage{multirow}       
\usepackage{makecell}       
\usepackage{array}          
\usepackage{tabularx}
\usepackage{todonotes}
\usepackage{xcolor}
\usepackage{comment}
\usepackage{algorithm}      
\usepackage{algpseudocode}  
\usepackage{xspace}

\usepackage{cleveref}

\usepackage{orcidlink}

\usepackage[utf8]{inputenc}
\DeclareUnicodeCharacter{2212}{-}

\newlength{\heightOfAvocado}

\newcommand{\nameNoPic}{{\sffamily AVOCADO}}
\newcommand{\nameTitle}{\nameNoPic~\includegraphics[height=0.4cm]{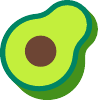}\xspace}
\newcommand{\name}{\nameNoPic~\settoheight{\heightOfAvocado}{A}\includegraphics[height=1.2\heightOfAvocado]{figures/avoc.pdf}\xspace}

\newcommand{\mypar}[1]{\smallskip\noindent\textbf{#1.}}

\definecolor{brainstorm}{rgb}{0.0,0.6,0.6}
\definecolor{first}{rgb}{0.85,0.5,0.5}
\definecolor{ok}{rgb}{0.4,0.3,0.3}
\definecolor{happy}{rgb}{0,0.0,0.0}

\begin{document}
\title{\nameTitle:\\ The Streaming Process Mining Challenge}


%
\titlerunning{\name: The Streaming Process Mining Challenge}

\author{
    Christian Imenkamp\inst{1} \orcidlink{0009-0007-4295-1268} \and
    Andrea Maldonado\inst{2} \orcidlink{0009-0009-8978-502X} \and
    Hendrik Reiter\inst{3} \orcidlink{0009−0003−8544−0012} \and\\
    Martin Werner\inst{2} \orcidlink{0000-0002-6951-8022} \and
    Wilhelm Hasselbring\inst{3} \orcidlink{0000-0001-6625-4335} \and
    Agnes Koschmider \inst{1} \orcidlink{0000-0001-8206-7636} \and
    Andrea Burattin\inst{4}\ \orcidlink{0000-0002-0837-0183}
}

\institute{University of Bayreuth, Bayreuth, Germany\\ \email{\{christian.imenkamp,agnes.koschmider\}@uni-bayreuth.de} \\ \and
School of Engineering and Design, Technical University of Munich, Germany 
\email{\{andrea.maldonado\, martin.werner\}@tum.de}\\ \and
Christian-Albrechts-University Kiel, Kiel, Germany\\ \email{\{hendrik.reiter,hasselbring\}@email.uni-kiel.de} \\ \and
DTU Compute, Technical University of Denmark, Kongens Lyngby, Denmark\\ \email{andbur@dtu.dk} \\}

\authorrunning{C. Imenkamp et al.}

\maketitle
\begin{abstract}
Streaming process mining deals with the real-time analysis of streaming data. 
Event streams require algorithms capable of processing data incrementally. 
To systematically address the complexities of this domain, we propose AVOCADO, a standardized challenge framework that provides clear structural divisions:
separating the concept and instantiation layers of challenges in streaming process mining for algorithm evaluation.
The \name evaluates algorithms on streaming-specific metrics like accuracy, Mean Absolute Error (MAE), Root Mean Square Error (RMSE), Processing Latency, and robustness.
This initiative seeks to foster innovation and community-driven discussions to advance the field of streaming process mining.
We present this framework as a foundation and invite the community to contribute to its evolution by suggesting new challenges, such as integrating metrics for system throughput and memory consumption, and expanding the scope to address real-world stream complexities like out-of-order event arrival.

\keywords{Streaming Process Mining  \and Streaming Conformance \and Concept Drifts \and Streaming Challenge.}
\end{abstract}

\section{Introduction}
\label{sec:intro}

Traditionally, process mining focuses on process discovery, the task of extracting process models from event data. These event logs, typically represent finite sets of recorded activities, often used to model the flow of a business process. While event logs provide a detailed record of past events, they are finite and often limited in the scope they capture, which can restrict the ability of process mining techniques to generalize. In recent years, the need to extend process mining techniques to handle event streams has grown and thus to efficiently handle continuous, potentially infinite sequences of events. Event streams better reflect real-time, dynamic business processes and offer a richer, more comprehensive data source for analysis.

The shift from event logs to event streams in process mining introduces several challenges that challenge process mining. While event logs consist of finite, static data representing past activities, event streams are continuous and real-time, requiring algorithms to process data as it arrives. This real-time processing demands algorithms that can incrementally update process models without access to the entire dataset at once, placing constraints on both memory and computational resources.
Beyond the new challenges, the real-time environment also introduces new opportunities for process mining. In offline processing, the discovery of process models from complete historical data is central. Conversely, real-time environments can emphasize tasks more suited to immediate actionability. In particular, real-time discovery offers limited practical value, as practitioners typically do not benefit from observing an evolving process model live. Conversely, streaming conformance checking becomes significantly more valuable in real-time scenarios. A low conformance score indicates an immediate deviation from expected behavior during the current process execution. 


To address these challenges, we propose \name, which is designed to evaluate process mining algorithms specifically for event streams. In particular, \name will focus on streaming conformance algorithms. It provides a standardized framework for assessing algorithms on their ability to process continuous event data while balancing accuracy and performance under drifting real-time conditions.
By using synthetic event data generated from complex process models, \name offers clear, objectified evaluation criteria that reflect the challenges inherent in working with event streams.
This challenge advances the state of process mining by driving the development of algorithms that can learn the expected behavior from event streams, while meeting the system and resource constraints typical of real-time, large-scale environments.
Ultimately, \name aims to foster process mining techniques allowing organizations to gain insights into their business processes in real-time, improving decision-making and operational efficiency. 

\begin{figure}
    \centering
    \includegraphics[width=0.7\linewidth]{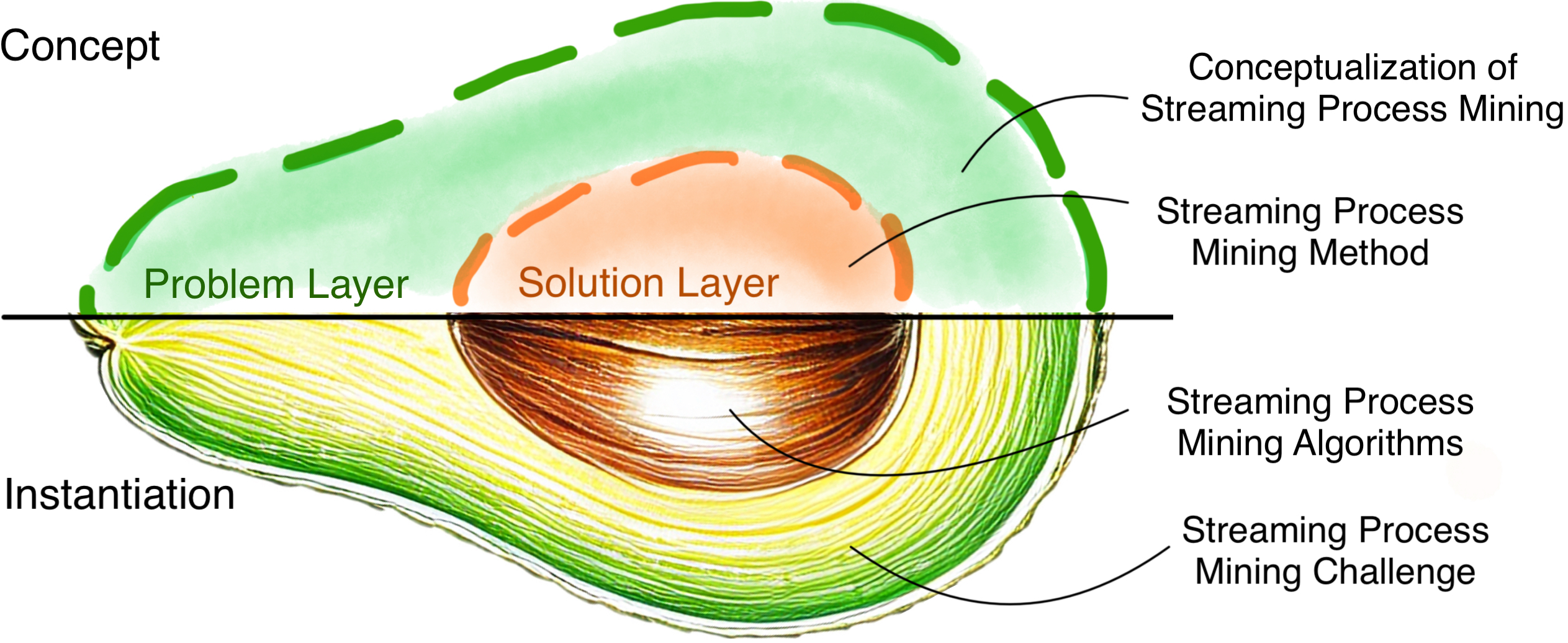}
    \caption{\name: Unveiling the problem's `pulp' to expose the solution set `seed', conceived in the formalization of SPM and embodied in this challenge.}
    \label{fig:layers}
\end{figure}

The conceptual framework guiding our work is visually depicted in Figure~\ref{fig:layers}. 
This \name representation structures the Streaming Process Mining domain into distinct layers. At the conceptual level, it differentiates between the \textit{problem layer} and the \textit{solution layer}. 
The instantiation of these concepts progresses from \textit{Streaming Process Mining Methods} to \textit{Streaming Process Mining Algorithms}, and the \textit{Conceptualization of Streaming Process Mining} and culminating in our \textit{Streaming Process Mining Challenge}.
This layered model illustrates how the proposed challenge serves as a concrete instantiation for evaluating algorithms within the broader context of streaming process mining.


\section{Related Work}
\label{sec:relatedWork}

This section discusses related work in terms of our addressed problem (see \Cref{subsec:problem_layer}) and the related literature to our solution (see \Cref{subsec:solution_layer}).

\subsection{Problem Layer}
\label{subsec:problem_layer}

\subsubsection{Process Discovery Contest}
The Process Discovery Contest (PDC), hosted annually at the International Conference on Process Mining (ICPM), serves as a prominent platform for evaluating the effectiveness and efficiency of process discovery algorithms.
Historically, the contest has been structured around the offline batch processing of static event logs \cite{carmona2017summary}.
The traditional PDC format, as outlined by \cite{carmona2017summary} and the ICPM 2025 contest descriptions\footnote{ICPM Process Discovery Contest 2025: \url{https://icpmconference.org/2025/process-discovery-contest/}}, involves:

\begin{itemize}
    \item \textbf{Static Event Logs}: Participants receive pre-generated, finite event logs (e.g., XES files) as input, clearly indicating the beginning and end of the data available for model discovery.
    \item \textbf{Batch Processing}: Algorithms are expected to process the entire log to produce a single, complete process model (e.g., in PNML or BPMN format).
    \item \textbf{Offline Evaluation}: Evaluation metrics such as fitness, precision, generalization, and simplicity are computed on the discovered model against a "ground truth" model derived from the static log. This process does not impose real-time constraints on continuous updates or performance.
    \item \textbf{Focus on Overfitting/Underfitting}: The primary objective is to achieve an optimal balance between overfitting (generating overly restrictive models) and underfitting (producing overly general models) to the provided log.
\end{itemize}

The traditional focus of process mining lies in analyzing historical static event logs for process discovery, monitoring, and improvement.
However, the increasing availability of real-time event data from various IT systems has driven a paradigm shift towards streaming process mining.
This emerging field addresses the unique challenges posed by unbounded, continuous streams of events, which are not present in conventional offline analysis.
The fundamental objective of streaming process discovery is to incrementally generate a sequence of process models, where each model is intended to reflect process behavior observed within a recent window of events, drawn from an infinite event stream. This inherent need for continuous model generation and adaptation fundamentally distinguishes streaming process mining from its offline counterparts, introducing the following requirements for a challenge. 
\begin{itemize}
    \item \textbf{Evaluation in Incomplete Information}: Streaming event logs are inherently incomplete, as traces are ongoing and future behavior is unknown, complicating discovery. Furthermore, they often contain noisy data due to errors or system glitches, which algorithms must robustly manage alongside concept drift, where process behavior changes over time \cite{batyuk_streaming_2020}. Additionally, processing events incrementally, often one by one, imposes strict resource constraints and unawareness of future events \cite{van_zelst_event_2018}.
    \item \textbf{Handling of Concept Drifts}: The underlying process may evolve over time. This means that the mined models must adapt to such changes, balancing the ingestion of new behavior to avoid quick fluctuations due to short-term changes \cite{zellner_2021_doa}.   
    \item \textbf{Online Evaluation}: Unlike batch-based evaluation using complete logs for a single final model, online evaluation in streaming process mining demands metrics for a sequence of evolving models. This is compounded by inherently incomplete trace information, complicating discovery and quality assessment of dynamic process behavior. \cite{burattin_streaming_2018}
    \item \textbf{Significance of Ground Truth in an Online Scenario}: In an online scenario, the concept of a static "ground truth" model for evaluation becomes ambiguous due to continuous process evolution (concept drift). Evaluation must thus assess how well models generalize from partial and dynamic behavior, potentially against an evolving reference or based on real-time deviation detection.
\end{itemize}

\subsection{Solution Layer}
\label{subsec:solution_layer}
Various research efforts have addressed streaming process mining, each with distinct capabilities and limitations, highlighting open research avenues.

\mypar{Algorithms for Concept Drift and Non-Stationary Processes}
Batyuk and Voityshyn \cite{batyuk_streaming_2018,batyuk_streaming_2020} proposed streaming discovery methods within Lambda architectures. In particular, it emphasizes operational support and recent event prioritization, but lacks explicit drift detection or noise management. Potoniec et al. \cite{potoniec_continuous_2022} developed an Online Miner for rapid drift response in causal nets, without fully addressing noise resilience or efficiency. Zellner et al. \cite{zellner_2021_doa} improved drift detection precision via dynamic outlier aggregation but offered limited adaptive modeling. Colombo Tosatto et al. \cite{colombo_tosatto_managing_2025} tackled out-of-order events in compliance monitoring through reliable prefix identification and confidence scores, critical for robust analyses. While effective individually, these approaches often isolate drift and noise handling from core discovery processes.

\mypar{Declarative and Low-Level Event Data Approaches}
Burattin et al. \cite{burattin_online_2015,burattin_uncovering_2023} explored declarative online discovery and change detection using DCR graphs, yet neglected computational efficiency and explicit noise strategies. Lefebure et al. \cite{lefebure_real-world_2023} focused on deriving high-level attributes from technical event streams to handle frequency and asynchrony, though impacts on model quality remain unclear. These approaches broaden streaming discovery scope but incompletely address practical implementation challenges.

\mypar{Online Conformance Checking}
Online conformance checking methods provide real-time process assessments. Van Zelst et al. \cite{van_zelst_online_2019} introduced incremental prefix-alignments balancing memory and approximation accuracy. Burattin et al. \cite{burattin_online_2018} presented an in-vivo deviation detection framework without predefined case start points. Nagy and Werner-Stark \cite{nagy_alignment_based_2022} extended alignment techniques with data perspectives for incomplete executions. Lee et al. \cite{lee_orientation_2021} developed an HMM-based approach ensuring constant-time event orientation. Practical implementations, such as CCaaS \cite{weber_ccaas_nodate}, demonstrate operational viability. Yet, comprehensive metrics and optimal resource-accuracy trade-offs remain open challenges.

Despite advances, current streaming process discovery approaches primarily address individual challenges like scalability or basic drift handling in isolation. Integrative solutions combining adaptive modeling, resource efficiency, diverse drift management, and resilience to incomplete or noisy data remain crucial.
\section{Conceptualization of Streaming Process Mining}%
\label{sec:method}

Streaming process mining, focuses on the analysis and discovery of process models from \textit{event streams} (i.e., unbounded, real-time sequences of events). 
Unlike traditional process mining, which operates on static, finite, and complete event logs, streaming process mining processes data incrementally and continuously, often under constraints related to memory, latency, and computational resources.

Formally, let $S = \langle \dots, e_1, e_2, e_3, \dots \rangle$ denote an infinite stream of events, where each $e_i = (a,c)$ contains at least an activity label $a$ and a case id $c$ -- also accessible via projection operators, i.e., $\pi_a(e_i)=a$ and $\pi_c(e_i)=c$. Then the goal of streaming process discovery is to incrementally generate a sequence of process models $\langle M_1, M_2,\dots \rangle$ such that each model $M_t$, discovered after the event with index $t$, reflects the behavior observed in the stream $S$ over the latest $\delta$ events, i.e., $S_{[t-\delta,t]}$.


In traditional process mining, the ``rediscovery problem''~\cite{10.1007/978-3-319-07734-5_6} refers to whether a control-flow discovery algorithm can discover a model $M$ that exactly reflects the behavior observed in a log $L$ from which it was derived. In the streaming context, considering the challenges mentioned above, this becomes more complex. The objective shifts from whether the evolving model correctly generalizes from partial and evolving behavior. Furthermore, unlike incremental process mining~\cite{Cattafi2010}, which updates models from growing logs, streaming process mining must update the model often, discarding earlier data due to memory constraints and changes becoming more relevant.

\begin{figure}[t]
    \centering
    \includegraphics[width=\linewidth]{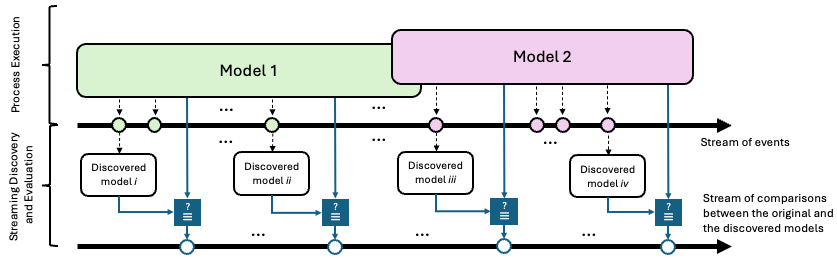}
    \caption{Conceptualization of the ``rediscovery problem'' in the context of Streaming Process Mining.}
    \label{fig:conceptualization}
\end{figure}
A graphical representation of the ``rediscovery problem'' is shown in Figure~\ref{fig:conceptualization}. Here, there are two models generating events. These two models follow one another with only a partial overlap. The events being generated into the ``stream of events'' refer to either one or the other. Each event is mined, and a corresponding model is discovered. Then, each discovered model must be compared to the original model from which the event was generated. In the picture, the result of such a comparison generates a stream of comparison values. 

As we aim for a modeling language-agnostic approach, the conceptualization of our rediscovery problem should not be bound to any fixed formalism (e.g., Petri net). Therefore, to accomplish this, rather than comparing the mined models with the original one, we decided to compute the conformance measure between the mined models and a log comprising the behavior of the reference model. It becomes clear that this evaluation simultaneously addresses two intertwined dimensions: the quality of the discovered model and the reliability of the conformance assessment itself. However, in the streaming context, conformance is not a straightforward measure, and it must be interpreted alongside other critical dimensions such as \textit{confidence} and \textit{completeness}. This, however, enables us to adopt a \textit{language-agnostic} approach.

Furthermore, we address the internal procedure and process of the challenge and the tests. For this, we provide \Cref{alg:algorithm_evaluation}: 
First (lines 1-4), two process models (i.e., $p$ and $k$) are constructed. While $p$ is fully random, $k$ contains at least 85\% of the same activities as $k$ ($k.a$ indicates the set of activities of $k$).  Furthermore, $p$ and $k$ have entirely distinct directly-follow relations ($k.df$ indicates the set of directly-follow relations of $k$), i.e., $k$ is a reordering of $p$. The union of $p$ and $k$ is called $w$ (i.e., a driving log). While $S$ denotes the stream in general, $S_{train}$ can be used by the algorithms to learn the behavior of the stream. Furthermore, $S_{val}$ is the subset of the stream to evaluate the algorithm. $S_{train}$ is constructed using a play-out of $p$, and is used for the warm-up phase where the algorithms can learn the behavior of the stream (lines 7-9). Next (lines 5-6), is the generation of the event stream for validation by construction of a drifting stream with $p \rightarrow w \rightarrow k$. Then, the ground truth ($gt$) is constructed:
Let denote $e \in S_{val}$ and $gt(e)$ the ground truth conformance of event e:
\[
gt(e) =
\sum_{e \in S_{\mathrm{val}}}
\begin{cases}
1 & \text{if } e \in p \\
0.5 & \text{if } e \in w \\
0 & \text{if } e \in k
\end{cases}
\]

After that (line 10-13), is the main challenge, where the participant's algorithm gets each event using the conformance method. The calculated conformance (i.e., the user prediction if the event is from $p$ or $k$) is stored and compared against the ground truth (line 14). $E_{global}$ is the average distance between the calculated conformance and the ground truth. 

\begin{algorithm}
\caption{Algorithm Evaluation Procedure}
\begin{algorithmic}[1]
    \State $p \gets \text{GetRandomProcess}()$
    \State $k \gets \text{GetRandomProcess}()\quad\text{with } \frac{|k.a \cap p.a|}{|k.a|} \geq 85\% \wedge k.df \neq p.df $
    \State $w \gets p \,\cup\, k$
    \State $\mathcal{S}_{\mathrm{train}} \gets \text{GenerateStream}(p)$
    \State $\mathcal{S}_{\mathrm{val}} \gets\; \text{GenerateStream}(p,100)\; \Vert\; \text{GenerateStream}(w,50)\; \Vert\; \text{GenerateStream}(k,100)$
    \State $\mathbf{gt} \gets \bigl[\underbrace{1,\dots,1}_{100},\,\underbrace{0.5,\dots,0.5}_{50},\,\underbrace{0,\dots,0}_{100}\bigr]$
    \vspace{1ex}
    \ForAll{$e \in \mathcal{S}_{\mathrm{train}}$}
        \State $a.\mathrm{learn}(e)$
    \EndFor
    \vspace{1ex}
    \State $\mathbf{c} \gets [\,]$
    \ForAll{$e \in \mathcal{S}_{\mathrm{val}}$}
        \State $Append\bigl(\,a.\mathrm{conformance}(e)\bigr)$ to $\mathbf{c}$
    \EndFor
    \vspace{1ex}
    \State $E_{\mathrm{global}} \;\gets\; \displaystyle\sum_{i=1}^{|\mathbf{c}|} \mathrm{compare}\bigl(\mathbf{gt}_i,\;\mathbf{c}_i\bigr)$
\end{algorithmic}
\label{alg:algorithm_evaluation}
\end{algorithm}


\section{Streaming Process Mining Challenge}
\label{sec:streaming_process_mining_challenge}

We now turn the discussion to \name. Our repository is publicly available\footnote{\url{https://github.com/chimenkamp/Streaming-Process-Mining-Challenge}} and the challenge can be accessed online.\footnote{\url{https://streaming-process-mining-challenge.onrender.com/}}
First, we address the requirements and general design of the algorithms in \Cref{subsec:algorithm_requirements}. Next, in \Cref{subsec:participation}, we present the underlying pipeline and the steps required to participate in the challenge. Finally, we discuss the parameters, metrics, and the structure of the event stream in \Cref{subsec:parameters}.  

\subsection{Algorithm Requirements}
\label{subsec:algorithm_requirements}

Participants must adhere to a few general guidelines when developing the conformance algorithm. This is necessary to ensure comparability, reproducibility, and successful execution during the challenge. 

First, algorithms must include a main class that implements the provided interface (i.e., \textit{BaseAlgorithm}). This is necessary to ensure that the required methods are implemented and aligned with the predefined structure. The interface includes two methods (\textit{learn} and \textit{conformance}). The \textit{learn} method is responsible for the \textbf{Warm-Up} phase, where the algorithm has the opportunity to learn the behavior of the stream. The \textit{conformance} method is called with each new event and should return the current conformance calculated by the algorithm. 

Second, participants must include a clear and executable start script located in the root of the directory. This script should contain the main class that implements the interface and all the necessary configurations. 
Additionally, submissions must not contain any top-level code that is executed by just loading the script (e.g., \texttt{if \_\_name\_\_ == "\_\_main\_\_"}). 
Moreover, algorithms must be capable of running solely on the CPU. We are explicitly excluding GPU support for the challenge. 
Although participants may utilize machine learning techniques, any required computation must occur within the CPU. 

Finally, participants must explicitly specify any external Python dependencies in a \textit{requirements.txt} file located in the root directory. Dependencies will be installed using the standard Python package manager (PIP) prior to the algorithm evaluation. 

\subsection{Participation}
\label{subsec:participation}

\begin{figure}
    \centering
    \includegraphics[width=0.9\linewidth]{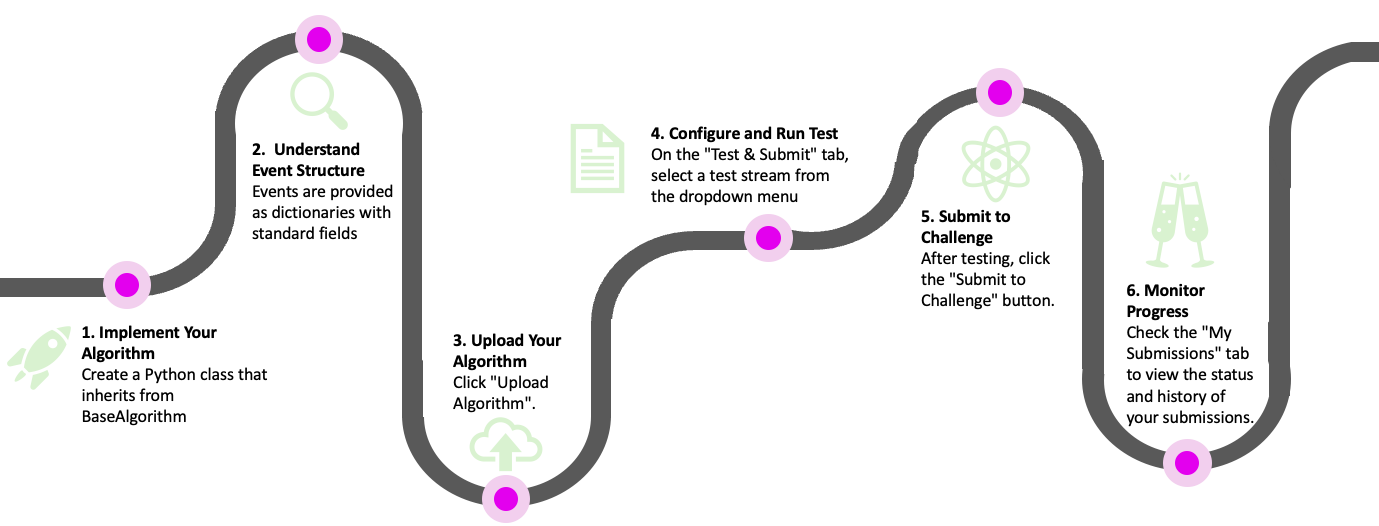}
    \vspace{-2em}
    \caption{High-Level overview of the steps required to participate in the challenge}
    \label{fig:pipeline}
\end{figure}

The challenge itself is web-based and can be accessed online\footnote{\url{https://streaming-process-mining-challenge.onrender.com/}}. To simplify participation in the challenge, we now address the necessary steps to submit an algorithm (see \Cref{fig:pipeline}). The participation steps are detailed below: 

\textbf{Implement Your Algorithm}. The implementation of the algorithm should adhere to the requirements outlined in \Cref{subsec:algorithm_requirements}

\textbf{Understand Event Structure}. Events are provided as dictionaries containing standard fields: \texttt{case:concept:name} (case identifier), \texttt{concept:name} (activity name), \texttt{time:timestamp} (event timestamp). Furthermore, in test streams, an additional field \texttt{concept:origin} (origin log for concept drifts) is provided. 

\textbf{Upload Your Algorithm}. Participants navigate to the \textbf{Test \& Submit} tab in the user interface. It is possible to upload a single Python file or a ZIP archive containing all required files (including the requirements.txt). The platform will automatically detect the class that implements the interface.  

\textbf{Configure and Run Test}. After the upload, participants can run tests before submitting to the challenge. In particular, they can select different test streams and review the results. The results contain a conformance plot, performance metrics, and a baseline comparison (see \Cref{subsec:parameters} for more details).

\textbf{Submit To Challenge}. Post-testing participants submit their algorithm using the \textbf{Submit to Challenge} button. They must fill out a submission form, including team name, contract email, algorithm name, and a brief description of their algorithm. 

Submissions' statuses and histories can be reviewed under the \textbf{My Submissions} tab. Additionally, the \textbf{Leaderboard} tab provides real-time rankings and performance comparisons among all participating teams.

\subsection{Parameters and Metrics}
\label{subsec:parameters}


Following, we address the evaluation metrics and parameters provided by the challenge: 

\textbf{Accuracy} The percentage of predictions within 10\% of the ground truth conformance score. 
\textbf{Mean Absolute Error (MAE)} Average absolute distance between the algorithm conformance scores and the baseline scores. 
\textbf{Root Mean Square Error (RMSE)} Squared root of the mean squared difference between the algorithm and the ground truth conformance scores. 
\textbf{Processing Latency} Normalized metric based on average processing time per event, favoring lower times.
\textbf{Robustness Score} Reflects the number of major errors (difference greater than 30\% from the ground truth).

These metrics are combined into the score displayed on the leaderboard. The \textit{score} is calculated by: 
\begin{dmath*}
    \mathit{score} = (0.3 \times \mathit{accuracy}) + (0.25 \times \mathit{MAE}) + (0.2 \times \mathit{RMSE}) + (0.15 \times \mathit{latency}) + (0.1 \times \mathit{robustness})
\end{dmath*}
The corresponding weights are open for discussion and may be adjusted on the basis of feedback from the community.

Furthermore, the platform includes several parameters for testing purposes. 
Participants can configure: 
\textbf{Test Stream Selection}: Choose from predefined streams (e.g., ``Sudden'', ``Gradual'', ``Incremental'', ``Recurring''), each described by drift characteristics and baseline behavior.
\textbf{Number of Cases}: Specify the event stream size by cases.


\section{Assessment of Requirements}\label{sec:results}

\begin{figure}
    \centering
    \includegraphics[width=0.85\linewidth]{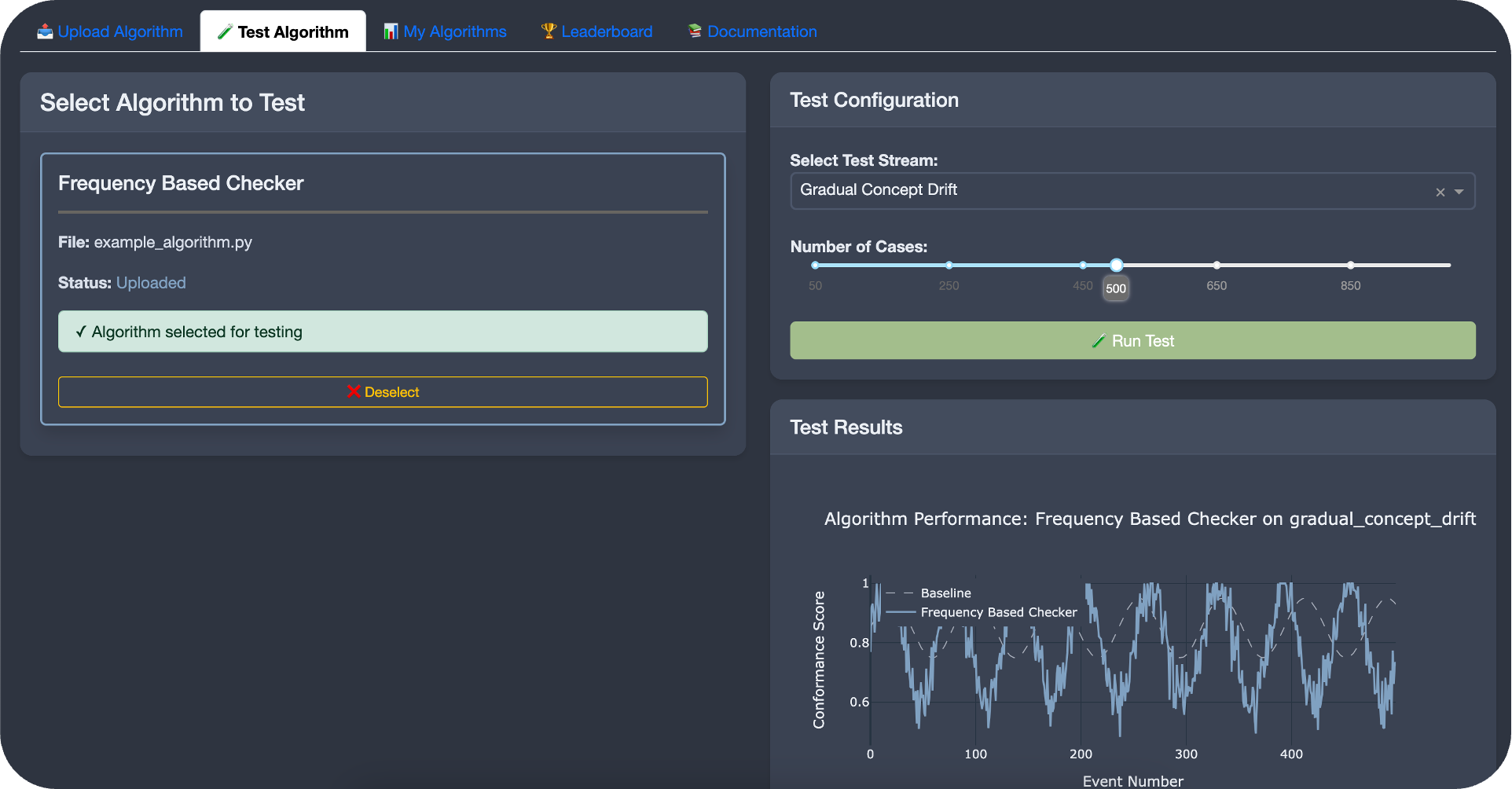}
    \vspace{-1em}
    \caption{Case-Study to assess the requirements of \name based on the frequency-based conformance algorithm}
    \label{fig:screenshot-challenge}
\end{figure}

We evaluate how the proposed \name addresses the requirements identified for streaming process mining (see \Cref{subsec:problem_layer}). In particular, we introduce a simple baseline algorithm as a test case to assess the requirements. It is called the \textit{Frequency-Based Conformance Algorithm} and learns frequencies of activities and directly follows relations during the learning phase. More frequent activities or transitions are considered to be more conformant. \Cref{fig:screenshot-challenge} shows the web interface on the \textit{Test \& Submit} page. 

\textbf{Evaluation in Incomplete Information} The \name addresses this by requiring algorithms to incrementally update their results when a new event arrives at the system. The stream is non-deterministic, and therefore future events are impossible to know. Thus, the challenge explicitly evaluates the robustness of the model to incomplete traces. 

\textbf{Handling of Concept Drifts} The generation of event streams directly addresses this requirement (see \Cref{alg:algorithm_evaluation}). In particular, by constructing validation streams with controlled concept drift scenarios (i.e., sudden, recurring, gradual, incremental) the challenge allows a detailed assessment of algorithms. Furthermore, algorithms are assessed on their ability to learn the behavior (i.e., conformance) of the stream and are evaluated based on the expected drop in conformance after the stream. Hence, algorithms aren't expected to show robustness against concept drifts. 

\textbf{Online Evaluation} \name implements an online evaluation framework. For that, it continuously evaluates algorithms with metrics that reflect their real-time performance. The adaptability and accuracy are continuously monitored using metrics such as accuracy, MAE, RMSE, and robustness. This ensures a realistic evaluation of algorithm performance in dynamic environments.

\textbf{Significance of Ground Truth in an Online Scenario} In streaming contexts, defining and maintaining ground truth is challenging. \name addresses this by reducing the complexity and defining the simplest possible ground truth. If the new event is from the process before the drift, we expect a conformance of one after a conformance of zero.

\section{Conclusion}\label{sec:conclusion}

In this paper, we have addressed the emerging challenges of streaming process mining. In particular, the relevance of conformance in real-time environments. We have outlined the significant differences and new quality dimensions introduced by streaming data (i.e., including handling incomplete information, concept drift, real-time evaluation, and ambiguity of ground truth). To foster advancements in this area, we proposed the \name, based on the process discovery contest,  offering a standardized web-based platform to assess algorithms' performance. For that, we consider metrics like accuracy, MAE, RMSE, Processing Latency, and robustness. 

This paper, however, does not propose a definitive or prescriptive solution.
Instead, it is intended as an invitation for discourse and collaboration, specially about parameters, evaluation metrics, and their weights in our score.
Recognizing the complexity and novelty of streaming process mining, our proposed challenge framework and evaluation criteria are meant as a foundation for further discussion.
We acknowledge that our approach and selected metrics might not capture all nuances relevant for streaming settings. Thus, we explicitly invite researchers and practitioners to critique, refine, and expand upon our concepts and evaluation methods.
By embracing a community-based approach, we believe we can collectively establish a more robust, comprehensive understanding of streaming process mining and continuously improve the challenge design.

\bibliographystyle{splncs04}
\bibliography{references}
\end{document}